\documentclass[12pt]{article}
\pdfoutput = 1
\usepackage{amsmath,amssymb,bm,epsfig,afterpage}
\usepackage{mathtools}
\usepackage{cite}
\usepackage{here}
\usepackage{color}
\usepackage{tabularx}
\usepackage{simplewick} 
\usepackage{bbm}

\usepackage{comment}

\renewcommand{\thefootnote}{\fnsymbol{footnote}}
\newcommand{\vev}[1]{{\langle{#1}\rangle}}

\newcommand{\abs}[1]{\left|{#1}\right|}

\newcommand{\eps}{\epsilon}

\newcommand{\hsg}{\hat{1}}
\newcommand{\hdb}{\hat{2}}
\newcommand{\htr}{\hat{3}}

\newcommand{\epth}[2]{\epsilon^{#1}\theta^{#2}} 
\newcommand{\order}[1]{\mathcal{O}\left({#1}\right)}

\newcommand{\pr}{\prime}

\allowdisplaybreaks[3]
\newcolumntype{Y}{&gt;{\centering\arraybackslash}X} 

\usepackage{hyperref}

\begin{document}

\begin{titlepage}

\begin{flushright}
 {\tt
CTPU-PTC-23-04 \\
EPHOU-23-006  
}
\end{flushright}

\vspace{1.2cm}
\begin{center}
{\Large
{\bf
Quark and lepton hierarchies from $S_4'$ modular flavor symmetry  
}
}
\vskip 2cm
Yoshihiko Abe$^{\ a}$~\footnote{yabe3@wisc.edu}, 
Tetsutaro Higaki$^{\ b}$~\footnote{thigaki@rk.phys.keio.ac.jp}, 
Junichiro Kawamura$^{\ b,c}$~\footnote{junkmura13@gmail.com}
and
Tatsuo Kobayashi$^{\ d}$~\footnote{kobayashi@particle.sci.hokudai.ac.jp}

\vskip 0.5cm

{\it $^a$
Department of Physics, University of Wisconsin-Madison, Madison, WI 53706, USA
}\\[3pt]

{\it $^b$
Department of Physics, Keio University, Yokohama, 223-8522, Japan
}\\[3pt]

{\it $^c$
Center for Theoretical Physics of the Universe, Institute for Basic Science (IBS),
Daejeon 34051, Korea
}\\[3pt]

{\it $^d$
Department of Physics, Hokkaido University, Sapporo 060-0810, Japan}\\[3pt]

\vskip 1.5cm

\begin{abstract}
We propose models in which the hierarchical structures of the masses and mixing 
in both quark and lepton sectors are explained by the $S_4^\prime$ modular flavor symmetry
near the fixed point $\tau \sim i\infty$.
The model provides the first explicit example which explains hierarchies of both quarks and leptons. 
The hierarchies are realized by powers of 
$\epsilon = e^{2\pi i \tau/4} = \mathcal{O}(0.01)$ and $2\,\mathrm{Im}\,\tau \sim 5$,  
where $\tau$ being the modulus. 
The small parameter $\eps$ plays a role of flavon in the Froggatt-Nielson mechanism 
under the residual $Z_4^T$ symmetry, 
and powers of $2\,\mathrm{Im}\,\tau$ in the Yukawa couplings are controlled by modular weights via the canonical normalization. 
The doublet quarks are identified to a $S_4^\pr$ triplet 
to explain the hierarchical structure of the quark mixing angles, 
while the doublet leptons are composed of three singlets 
for the large mixing angles in the lepton sector. 
We show that the $S_4^\prime$ modular symmetry alone can explain the hierarchies in both quark and lepton 
sectors by $\mathcal{O}(1)$ coefficients. 
\end{abstract}
\end{center}
\end{titlepage}

\clearpage

\renewcommand{\thefootnote}{\arabic{footnote}}
\setcounter{footnote}{0}

\newcommand{\la}{{\lambda}}
\newcommand{\ka}{{\kappa}}
\newcommand{\mQ}{{m^2_{\tilde{Q}}}}
\newcommand{\mU}{{m^2_{\tilde{u}}}}
\newcommand{\mD}{{m^2_{\tilde{d}}}}
\newcommand{\mL}{{m^2_{\tilde{L}}}}
\newcommand{\mE}{{m^2_{\tilde{e}}}}
\newcommand{\mhu}{{m^2_{H_u}}}
\newcommand{\mhd}{{m^2_{H_d}}}
\newcommand{\ms}{{m^2_S}} 
\newcommand{\Ala}{{A_\lambda}}
\newcommand{\Aka}{{A_\kappa}} 

\newcommand{\id}[1]{\mathbf{1}_{#1}}
\newcommand{\ol}[1]{\overline{#1}}
\newcommand{\Lcal}{\mathcal{L}}
\newcommand{\Mcal}{\mathcal{M}}
\newcommand{\Ncal}{\mathcal{N}}
\newcommand{\Ycal}{\mathcal{Y}}
\newcommand{\sg}{\sigma}
\newcommand{\sgb}{\overline{\sigma}}
\newcommand{\del}{\partial}

\newcommand{\htm}{\hat{m}}
\newcommand{\tU}{\widetilde{U}}
\newcommand{\SL}[2]{\mathrm{SL}({#1},{#2})}
\newcommand{\natN}{\mathbb{N}}
\newcommand{\intZ}{\mathbb{Z}}
\newcommand{\oGam}{\overline{\Gamma}} 
\newcommand{\Gam}{{\Gamma}} 
\newcommand{\Ita}{\mathrm{Im}\,\tau}

\newcommand{\CKM}{\mathrm{CKM}}
\newcommand{\SM}{\mathrm{SM}}
\newcommand{\hY}{\hat{Y}}

\newcommand{\nop}[1]{\textcolor{blue}{#1}}

\definecolor{darkviolet}{rgb}{0.58, 0.0, 0.83}
\newcommand{\violed}[1]{{\color{darkviolet}#1}}
\newcommand{\magenta}[1]{{\color{magenta}#1}}
\def\Re{\mathop{\mathrm{Re}}}
\def\Im{\mathop{\mathrm{Im}}}
\newcommand{\mr}{\mathrm}

\section{Introduction}
\label{sec-Intro} 

Modular flavor symmetry is an intriguing way to understand 
the origin of the flavor structure of the quarks and leptons in the Standard Model (SM)~\cite{Feruglio:2017spp,Kobayashi:2018vbk,Penedo:2018nmg,Novichkov:2018nkm,Ding:2019xna,Liu:2019khw,Novichkov:2020eep,Liu:2020akv,Liu:2020msy}. 
Yukawa coupling constants are transformed as modular forms under the modular symmetry, 
and hence these are holomorphic functions of the modulus $\tau$. 
The finite modular symmetries $\Gamma_N$, $N\in\natN$, are generalizations  
of the discrete non-Abelian flavor symmetries~\cite{deAdelhartToorop:2011re} 
which have been extensively studied in the literature~\cite{Altarelli:2010gt,Ishimori:2010au,Ishimori:2012zz,Hernandez:2012ra,King:2013eh,King:2014nza,Tanimoto:2015nfa,King:2017guk,Petcov:2017ggy,Feruglio:2019ybq,Kobayashi:2022moq}.  
For instance, $\Gamma_4$ is isomorphic to $S_4$ symmetry. 
Phenomenology of the modular flavor symmetries 
have been studied in Refs.~\cite{Criado:2018thu,Kobayashi:2018scp,Ding:2019zxk,Novichkov:2018ovf,Kobayashi:2019mna,Wang:2019ovr,Chen:2020udk,deMedeirosVarzielas:2019cyj,Asaka:2019vev,Asaka:2020tmo,deAnda:2018ecu,Kobayashi:2019rzp,Novichkov:2018yse,Kobayashi:2018wkl,Okada:2018yrn,Okada:2019uoy,Nomura:2019jxj,Okada:2019xqk,Nomura:2019yft,Nomura:2019lnr,Criado:2019tzk,King:2019vhv,Ding:2019gof,deMedeirosVarzielas:2020kji,Zhang:2019ngf,Nomura:2019xsb,Kobayashi:2019gtp,Lu:2019vgm,Wang:2019xbo,King:2020qaj,Abbas:2020qzc,Okada:2020oxh,Okada:2020dmb,Ding:2020yen,Okada:2020rjb,Okada:2020ukr,Nagao:2020azf,Wang:2020lxk,Okada:2020brs,Yao:2020qyy,Kuranaga:2021ujd}.

Recently, it has been shown that the modular flavor symmetry 
can explain the hierarchical structure of the quark masses 
and the Cabbibo-Kobayashi-Maskawa (CKM) mixing matrix, 
based on modular $A_4$~\cite{Petcov:2022fjf}, 
$\Gamma_6$~\cite{Kikuchi:2023cap}, $S_4^\pr$~\cite{Abe:2023ilq}  
and $A_4\times A_4\times A_4$~\cite{Kikuchi:2023jap} symmetries.  
The lepton sector is also studied in Refs.~\cite{Feruglio:2021dte,Novichkov:2021evw}. 
In these models, hierarchical structures are induced 
where the modulus $\tau$ has its vacuum expectation value (VEV) near a fixed point, 
so that a residual
$\mathbbm{Z}_n$ 
symmetry is approximately unbroken. 
The residual symmetry realizes
the well-known Froggatt-Nielsen (FN) mechanism~\cite{Froggatt:1978nt,Higaki:2019ojq}.  
For instance, 
the $m$-th power of flavon VEV is replaced by $e^{-2\pi m\Ita/N}$ 
when the modulus is stabilized near $\tau \sim i\infty$ 
and the symmetry generated by $T$, namely $\mathbbm{Z}_N^T$, is approximately unbroken. 
The residual symmetries are used for model buildings in Refs.~\cite{Novichkov:2018ovf,Novichkov:2018yse,Novichkov:2018nkm,Okada:2020brs,Novichkov:2021evw}.

In this work, we extend the analysis in Ref.~\cite{Abe:2023ilq} 
utilizing $\Gamma_4^\pr \simeq S_4^\pr$ symmetry to explain the quark hierarchy 
at $\tau \sim i\infty$.  
First of all, the lepton sector is studied under the $S_4^\pr$ symmetry, 
so that the mass hierarchies of the charged leptons 
are explained in the same manner as those of the quarks, 
as well as the neutrino oscillation data. 
This model provides the first explicit example 
in which the hierarchies in both quark and lepton sectors 
are explained by a common flavor modular symmetry.  
Secondly, we consider the co-existence of non-hatted and hatted representations 
in a same type of fermions, 
whereas we assumed that there is no such co-existence in Ref.~\cite{Abe:2023ilq}. 
This generalization allows us to explain the fermion hierarchies 
by completely $\order{1}$ coefficients 
unlike the models proposed in our previous study~\cite{Abe:2023ilq},  
where there are $\order{0.1}$ hierarchies among the coefficients~\footnote{
We pointed out that this small hierarchy can be interpreted 
by another modular $S_3$ flavor symmetry. 
}.

The rest of this paper is organized as follows. 
The modular flavor symmetry at $N=4$ is briefly reviewed in Sec.~\ref{sec-revmod}, 
and then the models are constructed in Sec.~\ref{sec-model}. 
Finally, we conclude in Sec.~\ref{sec-concl}.
The modular forms are shown in Appendix~\ref{app-S4p}.

\section{Modular symmetry at $N=4$} 
\label{sec-revmod}

We review the modular symmetry at level $N=4$. 
More detailed discussions are found in Ref.~\cite{Novichkov:2020eep}. 
We consider the series of groups $\Gamma(N)$, $N\in\natN$, 
called principal congruence subgroups, defined as 
\begin{align}
\Gamma(N) := \left\{
\begin{pmatrix}
          a & b \\
          c & d
\end{pmatrix} 
\in \SL{2}{\intZ},
\quad 
\begin{pmatrix}
    a & b \\
    c & d
\end{pmatrix} \equiv 
\begin{pmatrix}
1 & 0 \\
0 & 1
\end{pmatrix}
\mod N
\right\}, 
\end{align}  
where $\Gamma:= SL(2,\intZ) = \Gamma(1)$ 
is the special linear group of $2\times 2$ matrices of integers 
with determinant equal to one, i.e. $ad-bc=1$. 
The group $\Gamma$ acts on the complex variable $\tau$ ($\Ita>0$) as 
\begin{align}
 \tau \to \frac{a\tau +b}{c\tau+d}.    
\end{align}
Under the actions of these generators, $\tau$ is transformed as
\begin{align}
 \tau \xrightarrow{S} -\frac{1}{\tau}, 
\quad 
 \tau \xrightarrow{T} \tau+1, 
\quad 
 \tau \xrightarrow{R} \tau.    
\end{align} 
Since $R$ does not change $\tau$, one can consider the quotient group 
$\oGam:= \Gamma/\intZ_2^R$, 
where $\intZ_2^R$ being $\intZ_2$ symmetry generated by $R$, 
which has one-to-one correspondence with the action to $\tau$.    
The finite modular group $\Gamma_N^\pr$ ($\Gamma_N$)  
is defined as a quotient group $\Gamma_N^\pr := \Gamma/\Gamma(N)$ 
($\Gamma_N := \oGam/\Gamma(N)$). 
The group $\Gamma_N^\prime$ is generated by the generators, 
\begin{align}
 S = 
\begin{pmatrix}
 0 & 1 \\ -1 & 0 
\end{pmatrix}, 
\quad 
 T = 
\begin{pmatrix}
 1 & 1 \\ 0 & 1 
\end{pmatrix}, 
\quad 
 R = 
\begin{pmatrix}
 -1 & 0 \\ 0 & -1 
\end{pmatrix}, 
\end{align}
satisfying the following relations, 
\begin{align}
 S^2 = R, \quad (ST)^3 = R^2 = T^N = \mathbf{1}, \quad  TR=RT. 
\end{align}
Those for $\Gamma_N$ are given by setting $R=\mathbf{1}$. 
In a case of $N=4$, $\Gamma_4$ is isomorphic to the $S_4$ symmetry, 
and $\Gamma_4^\pr$ is isomorphic to the double covering of $S_4$, namely $S_4^\pr$. 
Under the $S_4^\pr$ symmetry, there are 10 irreducible representations, 
\begin{align}
 1,1^\pr, 2, 3, 3^\pr
\quad\text{and}\quad
 \hsg, \hsg^\pr, \hdb, \htr, \htr^\pr,  
\end{align}
where the non-hatted and hatted representations transform under $R$ 
trivially and non-trivially, respectively,  
i.e. $R r = r$ and $R \hat{r} = -\hat{r}$ for a representation $r$. 
We choose the basis in which $T$ is diagonal and $S$ is real. 
The representation matrices of the doublet $2$ and the triplet $3$ 
are respectively given by 
\begin{align}
\label{eq-rhoSTr2}
 \rho_S(2) = \frac{1}{2}
\begin{pmatrix}
 -1 & \sqrt{3} \\ \sqrt{3} & 1 
\end{pmatrix}, 
\quad 
\rho_T(2) = 
\begin{pmatrix}
 1 & 0 \\ 0 & -1 
\end{pmatrix}, 
\end{align}
and 
\begin{align}
\label{eq-rhoSTr3}
 \rho_S(3) = -\frac{1}{2}
\begin{pmatrix}
 0 & \sqrt{2} & \sqrt{2} \\ 
\sqrt{2} & -1 & 1 \\ 
\sqrt{2} & 1 & -1 \\ 
\end{pmatrix}, 
\quad 
\rho_T(3)= 
\begin{pmatrix}
 -1 & 0 & 0 \\ 0 & -i & 0 \\ 0 & 0 & i 
\end{pmatrix}. 
\end{align}
The primed or/and hatted representations are related as 
\begin{align}
 \rho_S(r) =&\ -\rho_S(r^\pr) = -i \rho_S(\hat{r}) =  i \rho_S(\hat{r}^\pr), \notag \\ 
 \rho_T(r) =&\ -\rho_T(r^\pr) =  i \rho_T(\hat{r}) = -i \rho_T(\hat{r}^\pr),  \\ \notag
\id{}= \rho_R(r) =&\  \rho_R(r^\pr) =  - \rho_R(\hat{r}) = -\rho_R(\hat{r}^\pr).
\end{align}

\begin{table}[tb]
 \center
\caption{\label{tab-wgt}
The number of representations of the modular forms at the weight $k \le 11$ 
in the $S_4^\pr$ modular symmetry.   
The representations for odd weights should be understood as the hatted ones. 
}
\vspace{0.5cm}
\begin{tabular}[t]{c|ccccccccccc} \hline 
 weight &  1 & 2 & 3 & 4 & 5 & 6 & 7 & 8 & 9 & 10 & 11 \\ \hline\hline
 $1$     & 0 & 0 & 0 & 1 & 0 & 1 & 0 & 1 & 1 & 1 & 0   \\
 $1^\pr$ & 0 & 0 & 1 & 0 & 0 & 1 & 1 & 0 & 1 & 1 & 1    \\ 
 $2$     & 0 & 1 & 0 & 1 & 1 & 1 & 1 & 2 & 1 & 2 & 2   \\ 
 $3$     & 1 & 0 & 1 & 1 & 2 & 1 & 2 & 2 & 3 & 2 & 3   \\ 
 $3^\pr$ & 0 & 1 & 1 & 1 & 1 & 2 & 2 & 2 & 2 & 3 & 3   \\ 
\hline 
\end{tabular}
\end{table}

The modular form $Y(\tau)$ with representation $r$ and weight $k$ 
transforms under the finite modular symmetry $\Gamma_4^\pr$ as 
\begin{align}
 Y^{(k)}_{r} (\tau) \to 
 (c\tau+d)^k \rho(r)  Y^{(k)}_{r} (\tau), 
\end{align} 
where $\rho(r)$ is a representation matrix.  
We assume that a matter field $f$, with representation $r_f$ and weight $k_f$, 
transforms in the same manner, 
\begin{align}
 f \to  (c\tau+d)^{k_f} \rho(r_f) f. 
\end{align}
There are $2k+1$ independent modular forms at a weight $k$. 
For $k=1$, there is a $\htr$ representation, 
\begin{align}
\label{eq-Y1}
 Y^{(1)}_{\htr}(\tau) = 
\begin{pmatrix}
 \sqrt{2}\eps(\tau) \theta(\tau) \\ \eps^2(\tau) \\ -\theta^2(\tau)
\end{pmatrix}.   
\end{align}
The functions $\theta$ and $\eps$ are defined as~\cite{Novichkov:2018nkm} 
\begin{align}
 \theta(\tau):= \frac{\eta(2\tau)^5}{\eta(\tau)^2 \eta(4\tau)^2}, 
\quad 
 \eps(\tau):= \frac{2\eta(4\tau)^5}{\eta(2\tau)}, 
\end{align}
where $\eta(\tau)$ is the Dedekind eta function. 
Their $q$-expansions are given by 
\begin{align}
 \theta(\tau) = 1+2\sum_{n=1}^\infty q^{n^2}, 
\quad 
 \eps(\tau) = 2q^{1/4} \sum_{n=0}^\infty q^{n(n+1)},  
\end{align}  
with $q := e^{2\pi i\tau}$. 
These functions are approximately $\theta(\tau) \simeq 1$ and $\eps(\tau)\sim 2q^{1/4} \ll 1$, 
for $2\pi\Ita\gg 1$, where the symmetry generated by $T$, 
namely $\intZ_4^T$, is a good symmetry. 
The modular forms with higher weights can be constructed 
from products of $Y^{(1)}_{\htr}$. 
The number of representations for the weights $k\le 11$ are shown in Table~\ref{tab-wgt}, 
and the explicit forms of the modular forms used in the models are listed 
in Appendix~\ref{app-S4p}. 
In this work, we assume that normalizations of the modular forms, 
which can not be determined from the modular symmetry,  
do not induce additional hierarcihal structures other than those from $\eps$.

The residual symmetry $\intZ_4^T$ realizes the FN mechanism.  
Under the $\intZ_4^T$ symmetry, 
$q^{1/4}$ plays the role of flavon whose charge is unity, 
because $q^{1/4} \xrightarrow{T} i q^{1/4}$. 
From Eqs.~\eqref{eq-rhoSTr2} and~\eqref{eq-rhoSTr3}, 
the hierarchical structures of the modular forms of $2$ and $3$ are read as 
\begin{align}
 Y_2 \sim 
\begin{pmatrix}
 1 \\ \eps^2
\end{pmatrix}, 
\quad 
 Y_3 \sim 
\begin{pmatrix}
 \eps^2 \\ \eps^3 \\ \eps 
\end{pmatrix}.  
\end{align}
The hierarchical structures of the other representations can be read in the same way.

\section{Models}
\label{sec-model}

\subsection{Hierarchical structures} 
\label{sec-hie} 

We aim to explain the hierarchical structure of the quarks and leptons in the SM 
by the $S_4^\pr$ modular flavor symmetry.   
In terms of $\eps \sim \order{0.01}$, 
the hierarchies of the fermion masses may be expressed as 
\begin{align}
\label{eq-texture} 
 (m_u, m_c, m_t) \sim&\ (\eps^3, \eps^2, 1), 
\quad 
 (m_d, m_s, m_b) \sim 
\eps^p (\eps^2, \eps, 1), \quad 
 (m_e, m_\mu, m_\tau) \sim 
\eps^p (\eps^2, \eps, 1),   
\end{align}
where $p=0,1$, and the CKM and PMNS matrices are given by 
\begin{align}
\label{eq-texmix} 
V_\CKM 
\sim&\  
\begin{pmatrix}
 1 & \eps & \eps^{2} \\
 \eps & 1 & \eps \\
 \eps^{2} &\eps & 1 
\end{pmatrix},   
\quad 
V_{\mathrm{PMNS}}
\sim 
\begin{pmatrix}
1 & 1 & 1  \\
1 & 1 & 1  \\ 
1 & 1 & 1 
\end{pmatrix}.   
\end{align}
In addition to these hierarchical structures, 
there are hierarchical structures from $2\Ita = -(4/\pi) \log \eps \sim 5$ 
due to the canonical normalization of matter kinetic terms depending on the modular weights as discussed later. 
The level $N=4$ is the minimal number to realize these texture 
with up to $\eps^3$ which may be necessary to explain the observed values 
with $\order{1}$ coefficients.

The texture in Eq.~\eqref{eq-texture} is realized 
if the representations of the quarks and leptons are 
\begin{align}
\label{eq-creps}
u^c =&\ 1\oplus 1\oplus \hsg^\pr, 
\quad  
d^c = 
\begin{cases}
1 \oplus 1 \oplus 1 \\  
\hsg^\pr \oplus \hsg^\pr \oplus \hsg^\pr  
\end{cases}, 
\quad 
Q = 3,
\quad 
e^c =  
\begin{cases}
3 \\ \htr^\pr   
\end{cases},  
\quad 
L = 
 1 \oplus 1 \oplus 1, 
\end{align}
where the top (bottom) case for $d^c$ and $e^c$ realizes $p=1$ ($p=0$). 
There are equivalent combinations which give the same Yukawa couplings, 
e.g.  $u^c = 1^\pr\oplus 1^\pr\oplus \hsg$ and $Q=3^\pr$ for the up Yukawa couplings.  
The triplets can be split into a doublet and 
a singlet, or three singlets, 
but we do not consider these possibilities 
because these are less predictive than the case of the triplets.  
We consider the different texture of the quarks from the model
proposed in our recent analysis~\cite{Abe:2023ilq}. 
In the current paper, 
we allow the co-existence of the hatted and non-hatted representations, 
in a same type of fermions 
as assigned in 
$u^c$.      
This generalization allows $V_{tb} \sim \order{1}$ even if $Q$ is a triplet, 
unlike the model without the co-existence.

\begin{table}[t]
 \centering
\caption{\label{tab-mat} 
Assignments of the fermions and Higgs doublets 
under $G_{\mathrm{EW}} := SU(2)_L\times U(1)_Y$, $S_4^\pr$ and the weight $k$. 
}
\begin{tabular}[t]{c|ccc|cc|cc} \hline 
     & $u^c_i$ & $d^c_i$ & $Q$ & $e^c$ & $L_i$ & $H_u$ & $H_d$ 
\\ \hline \hline 
$G_{\mathrm{EW}}$ & $1_{-2/3}$ & {$1_{1/3}$} & {$2_{1/6}$} 
                  & $1_{1}$ & {$2_{-1/2}$}   
                  & $2_{1/2}$ & $2_{-1/2}$  \\ \hline
$S_4^\pr$  & $(1,1,\hsg^\pr)$ & $(1,1,1)$ or $(\hsg^\pr,\hsg^\pr,\hsg^\pr)$ & $3$ & $3$ or $\htr^\pr$
           & $(1,1,1)$
           & $1$ & $1$ \\ 
$k$        & $-k_{u_i}$ & $-k_{d_i}$ & $-k_{Q}$ & $-k_{e}$ & $-k_{L_i}$  
           & $0$ & 0 \\ \hline  
\end{tabular}
\end{table}

We study the models with representations of the quarks and leptons 
shown in Eq.~\eqref{eq-creps}, 
so that the texture in Eqs.~\eqref{eq-texture} and ~\eqref{eq-texmix} is realized.  
The matter contents of the quarks and leptons 
and our notation of the modular weights are summarized in Table~\ref{tab-mat}. 
The Yukawa coupling terms in the superpotential are schematically given by
\begin{align}
 W =&\ H_u \left\{ \sum_{a=1}^2 \alpha_a  \left( Q Y_3^{({k_{u_a}+k_Q})} \right)_1  u_a^c  
          + \alpha_3 \left( Q Y_{\htr}^{(k_{u_3}+k_Q)} u_3^c \right)_1 
\right\}  \\ \notag 
        & \quad 
          + H_d \sum_{i=1}^3 
 \left\{ \beta_{i}  \left( QY^{(k_{d_i}+k_{Q})}_{\mathbf{3}} d^c_i \right)_1   
        +\gamma_{i} L_i \left(  Y^{(k_{e}+k_{L_i})}_{\mathbf{3}} e^c \right)_{1}  
\right\} 
+ \sum_{i,j=1}^3 \frac{c_{ij}}{\Lambda} 
Y_{{1}}^{(k_{L_i} + k_{L_j})}   L_i H_u L_j H_u 
\\ \notag 
=:&\ H_u Q Y_u u^c + H_d Q Y_d d^c + H_d L Y_e e^c + H_u L^T C_n  H_u L,            
\end{align}
where $\mathbf{3} = 3$~($\htr$) for $p=1$ ($p=0$) in the Yukawa terms to which $H_d$ couples.  
Here, $(\cdots)_{1}$ 
is the trivial singlet combination of the product inside the parenthesis. 
The number of the coefficients $\alpha_i$, $\beta_i$ and $\gamma_i$ 
will change accordingly to the number of modular forms 
for a given representation $r$ and modular weight $k$, 
as explicitly shown after assigning modular weights. 
We consider the Weinberg operator for the Majorana neutrino masses 
given by the last term. 
We assume that $L$ is a non-hatted singlet 
such that the corresponding modular form is a trivial-singlet. 
The modular forms in the Weinberg operator are the non-trivial singlets $1^\pr$ 
if $L$ are the hatted singlets, 
but we do not consider this possibility for simplicity.

The K\"{a}hler potential of the chiral superfield $f$ with a weight $-k_f$ is given by  
\begin{align}
K\supset \frac{f^\dagger f}{(-i\tau+i\ol{\tau})^{k_f}},     
\end{align}
and hence the couplings are modified by the canonical normalization as 
\begin{align}
\label{eq-cnorm}
 \left[Y_f\right]_{ij} \to\left(\sqrt{2\,\Ita}\right)^{k_Y}  
        \left[Y_f\right]_{ij}, 
\quad 
\left[C_n\right]_{ij}\to\left(\sqrt{2\,\Ita}\right)^{k_{L_i}+k_{L_j}}\left[C_n\right]_{ij},
\end{align}
for $i,j = 1,2,3$ and $f=u,d,e$. Here, $k_Y$ is the weight of the Yukawa coupling.

The hierarchical structures of the Yukawa matrices before the canonical normalization 
are given by 
\begin{align}
\label{eq-hieY}
 Y_u  \sim 
\begin{pmatrix}
\eps^2 & \eps^2 &  \eps \\
\eps   & \eps   & 1  \\ 
\eps^3 & \eps^3 &  \eps^2 \\
\end{pmatrix}, 
\quad 
 Y_d  \sim  Y_e^T \sim 
\begin{pmatrix}
\eps^2 & \eps^2 & \eps^2 \\ 
\eps   & \eps   & \eps   \\ 
\eps^3 & \eps^3 & \eps^3 \\ 
\end{pmatrix}, 
\quad 
C_n \sim \frac{1}{\Lambda}
\begin{pmatrix}
1 & 1 & 1 \\ 
1 & 1 & 1 \\ 
1 & 1 & 1  
\end{pmatrix}.   
\end{align}
These structures realize hierarchical structures 
in Eqs.~\eqref{eq-texture} and ~\eqref{eq-texmix}.     
The CP phase from the modulus VEV appears only in $\eps(\tau)$ up to $\order{\eps^4}$, 
but the phases can be absorbed by the redefinition of the fermions. 
Thus there should be CP phases from $\order{1}$ coefficients 
rather than that from $\tau$. 
This fact also implies that the Yukawa couplings are independent of 
$\mathrm{Re}\,\tau$ as a good approximation.

\subsection{Model for $p=1$}

\begin{table}[t]
\centering 
\caption{\label{tab-obs}
Values of the masses and mixing angles at the fitted points in the models
in the case of $p = 1$ (left) and $p = 0$ (right). 
The second, third and fourth columns show the model predictions, 
experimental central values and its errors, respectively. 
See the text in detail.
$s_{ij}^Q$ is the mixing angle between the $i$-th and $j$-th quark in the CKM matrix 
and $s_{ij}^2$ is the square of mixing angle in the PMNS matrix. 
$\delta_{\mathrm{CKM}}$ and $\delta_{\mathrm{PMNS}}$ are the CP phases. 
$R^{21}_{32} := (m_{2}^2 -m_1^2)/(m_3^2-m_2^2)$ is the ratio of the neutrino mass squared difference.
These are defined in the standard parametrization~\cite{Workman:2022ynf}.  
} 
\begin{minipage}[t]{0.48\textwidth}
\begin{tabular}[t]{c|ccc} \hline 
obs. & value & center & error \\  \hline \hline 
$y_u$$/10^{-6}$ & 4.44 & 2.85 & 0.88 \\  
$y_c$$/10^{-3}$ & 1.481 & 1.479 & 0.052 \\  
$y_t$ & 0.5322 & 0.5320 & 0.0053 \\  \hline 
$y_d$$/10^{-5}$ & 1.94 & 1.93 & 0.21 \\  
$y_s$$/10^{-4}$ & 3.88 & 3.82 & 0.21 \\  
$y_b$$/10^{-2}$ & 2.097 & 2.100 & 0.021 \\  \hline 
$y_e$$/10^{-6}$ & 7.816 & 7.816 & 0.047 \\  
$y_\mu$$/10^{-3}$ & 1.6496 & 1.6500 & 0.0099 \\  
$y_\tau$$/10^{-2}$ & 2.808 & 2.805 & 0.028 \\  \hline\hline 
$s_{12}^Q$ & 0.22520 & 0.22541 & 0.00072 \\  
$s_{23}^Q$$/10^{-2}$ & 4.007 & 3.998 & 0.064 \\  
$s_{13}^Q$$/10^{-3}$ & 3.43 & 3.48 & 0.13 \\  
$\delta_{\mathrm{CKM}}$ & 1.2395 & 1.2080 & 0.0540 \\  \hline 
$R^{21}_{32}$$/10^{-2}$ & 3.053 & 3.070 & 0.084 \\  
$s_{12}^2$ & 0.302 & 0.307 & 0.013 \\  
$s_{23}^2$ & 0.547 & 0.546 & 0.021 \\  
$s_{13}^2$$/10^{-2}$ & 2.203 & 2.200 & 0.070 \\  
$\delta_{\mathrm{PMNS}}$ & -0.85 & -2.01 & 0.63 \\  \hline 
\end{tabular} 
\end{minipage}
\begin{minipage}[t]{0.48\textwidth}
\begin{tabular}[t]{c|ccc} \hline 
obs. & value & center & error \\  \hline \hline 
$y_u$$/10^{-6}$ & 3.27 & 2.74 & 0.85 \\  
$y_c$$/10^{-3}$ & 1.418 & 1.419 & 0.050 \\  
$y_t$ & 0.5030 & 0.5029 & 0.0050 \\  \hline 
$y_d$$/10^{-5}$ & 6.93 & 7.32 & 0.81 \\  
$y_s$$/10^{-3}$ & 1.383 & 1.450 & 0.078 \\  
$y_b$$/10^{-2}$ & 7.997 & 7.976 & 0.080 \\  \hline 
$y_e$$/10^{-5}$ & 2.966 & 2.966 & 0.018 \\  
$y_\mu$$/10^{-3}$ & 6.259 & 6.261 & 0.038 \\  
$y_\tau$ & 0.1077 & 0.1074 & 0.0011 \\  \hline\hline  
$s_{12}^Q$ & 0.22533 & 0.22541 & 0.00072 \\  
$s_{23}^Q$$/10^{-2}$ & 4.003 & 4.011 & 0.064 \\  
$s_{13}^Q$$/10^{-3}$ & 3.47 & 3.49 & 0.13 \\  
$\delta_{\mathrm{CKM}}$ & 1.2032 & 1.2080 & 0.0540 \\  \hline 
$R^{21}_{32}$$/10^{-2}$ & 3.070 & 3.070 & 0.084 \\  
$s_{12}^2$ & 0.307 & 0.307 & 0.013 \\  
$s_{23}^2$ & 0.541 & 0.546 & 0.021 \\  
$s_{13}^2$$/10^{-2}$ & 2.199 & 2.200 & 0.070 \\  
$\delta_{\mathrm{PMNS}}$ & -2.30 & -2.01 & 0.63 \\  \hline 
\end{tabular} 
\end{minipage}
\end{table}

Now we consider the model for $p=1$. 
The modular weights are given by 
\begin{align}
\label{eq-kI}
 (k_{u_1}, k_{u_2}, k_{u_3}) = (0,4,3), 
\quad 
 (k_{d_1}, k_{d_2}, k_{d_3}) = (0,2,4), 
\quad 
 k_Q = 4, 
\quad 
 k_e = 4,  
\quad 
 k_L = (0,2,4),  
\end{align}
so that the superpotential is given by 
\begin{align}
 W =&\ H_u \left\{\alpha_1  \left( Q Y_3^{(4)} \right)_1  u_1^c  
                + \alpha_2^{i_Y} \left( Q Y_3^{i_Y(8)} \right)_1  u_2^c    
                + \alpha_3^{i_Y} \left( Q Y_{\htr}^{i_Y(7)} u_3^c \right)_1 
\right\}  \\ \notag 
        &+ H_d 
  \left\{ 
     \sum_{a=1}^2 \beta_{a}  \left( QY^{(2+2a)}_{{3}}  \right)_1 d^c_a  
                + \beta_{3}^{i_Y}  \left( QY^{i_Y (8)}_{{3}}  \right)_1 d^c_3  \right\} 
   \\ \notag 
  &+ H_d \left\{
    \sum_{a=1}^2 \gamma_{a} L_a \left(  Y^{(2+2a)}_{{3}} e^c \right)_{1}  
               + \gamma_{3}^{i_Y} L_3 \left(  Y^{i_Y (8)}_{{3}} e^c \right)_{1}  
\right\}  
+ \sum_{i,j=1}^3 \frac{c_{ij}}{\Lambda} 
    Y^{(2i+2j-4)}_{{1}}  L_i H_u L_j H_u, 
\end{align}
where the summation over the modular forms $i_Y = 1,2$ are implicit. 
The modular weights of $d^c_i$, as well as  $L_i$ should be different 
among the flavors to make all of the fermions massive. 
For instance, the Yukawa matrix $Y_d$ has only one non-zero singular value 
if all of the modular weights are $4$ or $6$, 
where there is only one $3$ representation. 
Hence, the assignment in Eq.~\eqref{eq-kI} is the minimal number 
to have three massive down quarks and charged leptons. 
We assume that the 18 coefficients are real except $\alpha_3^1$ and $\gamma_3^1$ 
to realize the non-zero CP phases in the CKM and PMNS matrices. 
Note that $c_{12}$ for Weinberg operator is irrelevant, 
because there is no trivial singlet at $k=2$.  
Since the absolute size of the neutrino mass 
can be fitted by $\Lambda$, 
we only fit to the ratio of the neutrino mass squared difference 
$R^{21}_{32} := (m_{2}^2 -m_1^2)/(m_3^2-m_2^2)$.

We fitted $\Ita$, $\tan\beta := \vev{H_u^0}/\vev{H_d^0}$ and the coefficients to explain 
the Yukawa couplings and the CKM angles at the GUT scale in the MSSM 
with $M_{\mathrm{SUSY}} = 10~\mathrm{TeV}$ 
and neglecting the threshold corrections~\cite{Antusch:2013jca}.  
We assume the $1\%$ relative uncertainties for the third generation fermions, 
and those shown in Table~3 of Ref.~\cite{Antusch:2013jca} 
are used for the other observables.
The neutrino data is taken from Ref.~\cite{Workman:2022ynf} 
under the assumption of the normal ordering, 
as $m_1 < m_2 < m_3$ is naturally predicted from the weight assignment.

With the weight assignment, 
the hierarchical structures with the powers of $t:= 2\Ita$ is given by 
\begin{align}
\label{eq-estI}
 (y_u, y_c, y_t) \sim&\ \left(\eps^3/t^{3/2}, \eps^2 t^{1/2}, 1\right)y_t 
                 \sim (5\times 10^{-7}, 7\times 10^{-4}, 0.5), \\ \notag
 (y_d, y_s, y_b) 
 \sim 
 (y_e, y_\mu, y_\tau)
\sim&\ \left(\eps^3/t^{3/2}, \eps^2/t^{1/2}, \eps t^{1/2} \right) y_t 
\sim (5\times 10^{-7}, 1\times 10^{-4}, 0.03),      \\ \notag 
 (s_{12}^Q, s_{23}^Q, s_{13}^Q) \sim&\ \left(\eps, \eps, \eps^2 \right) 
\sim (0.02, 0.02, 0.0006),  \\ \notag  
 (R^{21}_{32}, s_{12}^2, s_{23}^2, s_{13}^2) \sim&\ 
\left(1/t^4, 1/t^2, 1/t^2, 1/t^4\right) \sim 
\left( 0.001, 0.03, 0.03, 0.001 \right), 
\end{align}
where $\Ita = 2.83$ and $y_t = 0.5$ are used for the numerical estimation.  
Here, $y_f$, where $f$ being the SM charged fermions, are the Yukawa couplings 
in the mass basis. 
Since the top Yukawa coupling $y_t$ is roughly given by $\alpha_3^1 t^{7/2}$, 
the coefficient $\alpha_3^1$ as well as the other coefficients 
should be so small that the factors from powers of $t$ are canceled.  
$s_{ij}^Q$ is the mixing angle between the $i$-th and $j$-th quark in the CKM matrix 
and $s_{ij}^2$ is the square of mixing angle in the PMNS matrix
in the standard parametrization~\cite{Workman:2022ynf}. 
Among the observables, some of the light fermion 
Yukawa couplings
$y_d$, $y_e$, $y_\mu$,  
the Cabbibo angle $s_{12}^Q$, 
and the neutrino mixing angles are smaller than the observed values by $\order{10}$.  
Most of the differences are explained by the modular forms, 
e.g. the factor $7\sqrt{2}$ in the coefficients of the element with $\order{\eps^3}$ in $Y^{2(7)}_{\htr}$. 
Whereas, $\abs{c_{13}/c_{33}} \sim t$ will be necessary 
to compensate the suppression by $\order{t}$ in the neutrino mixing.  
Since all of the values are predicted to be smaller than the observed values, 
there will be no cancellation among the different contributions.  
Thus, in the current model, 
the cancellation required to explain the charm mass in the model studied 
in Ref.~\cite{Abe:2023ilq} is absent.

We found the parameter sets which can explain the observed data within $1.8\sigma$. 
At the benchmark point, 
$\tan\beta = 3.6666$,  
$\Ita = 2.8258$, 
$|\alpha_3^1| = 1.2334\times 10^{-3}$, 
\begin{align} 
\frac{1}{|\alpha_3^1|} 
\begin{pmatrix} 
\alpha_{1} \\ \alpha^1_2 \\ \alpha^2_2 \\ \alpha^1_3 \\ \alpha^2_3 
\end{pmatrix} 
=&\ 
\begin{pmatrix} 
-0.6863 \\ -1.7953 \\ 0.8360 \\ e^{-1.6360i} \\ -1.1016 
\end{pmatrix} 
, \quad 
\frac{1}{|\alpha_3^1|} 
\begin{pmatrix} 
\beta_{1} \\ \beta_{2} \\ \beta^1_3 \\ \beta^2_3 
\end{pmatrix} 
= 
\begin{pmatrix} 
1.6470 \\ -1.8393 \\ 0.6233 \\ -0.6245 
\end{pmatrix} 
, \\ \notag 
\frac{1}{|\alpha_3^1|} 
\begin{pmatrix} 
\gamma_{1} \\ \gamma_2 \\ \gamma^1_3 \\ \gamma^2_3 
\end{pmatrix} 
=&\ 
\begin{pmatrix} 
-1.6692 \\ -2.7814 \\ 0.8032\times e^{0.9284i} \\ 2.7497 
\end{pmatrix} 
, \quad 
\frac{1}{|\alpha_3^1|} 
\begin{pmatrix} 
c_{11} \\ c_{22} \\ c_{33} \\ c_{13} \\ c_{23} 
\end{pmatrix} 
= 
\begin{pmatrix} 
1.7273 \\ -1.9311 \\ -0.6692 \\ -5.3871 \\ -2.0488 
\end{pmatrix}. 
\end{align} 
The absolute values of the coefficients are in the range $[0.62,5.4]$. 
As expected from the estimation Eq.~\eqref{eq-estI}, 
$\abs{c_{13}}$ is larger than the other coefficients 
to realize the neutrino mixing angles.  
This factor would be explained by numerical factors of modular forms 
of non-singlet representations when we consider the see-saw mechanism 
as a UV completion of the Weinberg operator.

\subsection{Model for $p=0$}

Now we consider the model with $p=0$. 
The modular weights are given by 
\begin{align}
 (k_{u_1}, k_{u_2}, k_{u_3}) = (4,8,7), 
\quad 
 (k_{d_1}, k_{d_2}, k_{d_3}) = (1,3,5), 
\quad 
 k_Q = 0, 
\quad 
 k_e = 1,  
\quad 
 k_L = (2,4,4),  
\end{align}
so that the superpotential is given by 
\begin{align}
 W =&\ H_u \left\{\alpha_1  \left( Q Y_3^{(4)} \right)_1  u_1^c  
                + \alpha_2^{i_Y} \left( Q Y_3^{i_Y(8)} \right)_1  u_2^c    
                + \alpha_3^{i_Y} \left( Q Y_{\htr}^{i_Y(7)} u_3^c \right)_1 
\right\}  \\ \notag 
        &+ H_d 
  \left\{ 
     \sum_{a=1}^2 \beta_{a}  \left( QY^{(1+2a)}_{{\htr}} d^c_a  \right)_1  
                + \beta_{3}^{i_Y}  \left( QY^{i_Y (5)}_{{\htr}}  d^c_3  \right)_1 \right\} 
   \\ \notag 
  &+ H_d \left\{\gamma_{1} L_1 \left(  Y^{(3)}_{\htr} e^c \right)_{1}  
     + \sum_{b=2}^3 \gamma_{b}^{i_Y} L_b \left(  Y^{i_Y (5)}_{\htr} e^c \right)_{1}  
\right\}  
+ \sum_{i,j=1}^3 \frac{c_{ij}}{\Lambda} 
    Y^{(8-2\delta_{i1}-2\delta_{j1})}_{{1}}  L_i H_u L_j H_u. 
\end{align}
We assigned the same modular weights for $L_2$ and $L_3$, 
so that there is no hierarchy in the second and third generations 
in the neutrino sector. 
The three charged leptons are still massive because there are two modular forms 
of the $\htr$ representation at $k=5$. 
There are 20 coefficients in the model. 
The hierarchical structure of the observables are estimated as 
\begin{align}
\label{eq-estII}
 (y_u, y_c, y_t) \sim&\ \left(\eps^3/t^{3/2}, \eps^2 t^{1/2}, 1\right) y_t 
\sim (1\times 10^{-7}, 3\times 10^{-4}, 0.5), \\ \notag
 (y_d, y_s, y_b) \sim&\ \left(\eps^2/t^3, \eps/t^2, 1/t \right) y_t
\sim (6\times 10^{-7}, 2\times 10^{-4}, 0.08),         \\ \notag 
 (y_e, y_\mu, y_\tau) \sim&\ \left( \eps^2/t^2, \eps/t, 1/t   \right) y_t
\sim (3\times 10^{-6}, 0.001, 0.08),  \\ \notag
 (s_{12}^Q, s_{23}^Q, s_{13}^Q) \sim&\ \left(\eps, \eps, \eps^2 \right) 
\sim (0.02, 0.02, 3\times 10^{-4}),  \\ \notag 
 (R^{21}_{32}, s_{12}^2, s_{23}^2, s_{13}^2) \sim&\ 
 \left( 1, 1/t^2, 1, 1/t^2 \right)\sim (1,0.03, 1, 0.03),  
\end{align}
where $\Ita = 3.07$. 
In this case, $y_u$, $y_d$, $s_{12}^Q$, $s_{13}^Q$ and $s_{12}^2$ 
are more than 10 times smaller than the observed values, 
especially $y_d$ is about $125$ times smaller. 
Again, the gaps in the quark sector can be explained by the modular forms, 
e.g. the factors of $5$ in the $\order{\eps^2}$ elements 
in $Y^{(3)}_{\htr}$ and $Y^{2(5)}_{\htr}$.  
Thus the observed values will be explained by the $\order{1}$ coefficients.
We assume that the neutrinos are in the normal ordering, 
because the model predicts $m_1 \lesssim m_2, m_3$ 
and $s_{12}^2, s_{13}^2 \lesssim s_{23}^2$ which more easily fits 
to it rather than the inverse ordering case.

We found the benchmark point which can explain the experimental values 
within $0.9\sigma$.
The values are given by 
$\tan\beta = 14.1755$,  
$\Ita = 3.0744$, 
$|\alpha_3^1| = 8.6886\times 10^{-4}$, 
\begin{align} 
\frac{1}{|\alpha_3^1|} 
\begin{pmatrix} 
\alpha_{1} \\ \alpha^1_2 \\ \alpha^2_2 \\ \alpha^1_3 \\ \alpha^2_3 
\end{pmatrix} 
=&\ 
\begin{pmatrix} 
-1.0336 \\ 1.2757 \\ -2.2480 \\ e^{0.5391i} \\ 4.8979 
\end{pmatrix} 
, \quad 
\frac{1}{|\alpha_3^1|} 
\begin{pmatrix} 
\beta_{1} \\ \beta_{2} \\ \beta^1_3 \\ \beta^2_3 
\end{pmatrix} 
= 
\begin{pmatrix} 
2.5696 \\ -3.3719 \\ 4.0290 \\ -0.8087 
\end{pmatrix} 
, \\ \notag 
\frac{1}{|\alpha_3^1|} 
\begin{pmatrix} 
\gamma_{1} \\ \gamma^1_2 \\ \gamma^2_2 \\ \gamma^1_3 \\ \gamma^2_3 
\end{pmatrix} 
=&\ 
\begin{pmatrix} 
4.0267 \\ 4.6090 \\ 0.8186 \\ 4.4630\times e^{-2.9326i} \\ -0.8028 
\end{pmatrix} 
, \quad 
\frac{1}{|\alpha_3^1|} 
\begin{pmatrix} 
c_{11} \\ c_{22} \\ c_{33} \\ c_{12} \\ c_{13} \\ c_{23} 
\end{pmatrix} 
= 
\begin{pmatrix} 
-1.3694 \\ 1.2175 \\ 1.3739 \\ 1.4887 \\ -6.0655 \\ -1.1843 
\end{pmatrix}.  
\end{align} 
The absolute values of the coefficients are within the range $[0.8, 6.1]$, 
and hence the ratio of the smallest and largest coefficients is $7.6$. 
The coefficient $c_{13}$ is $\order{t}$ larger than the other coefficients 
of the Weinberg operator to obtain the large mixing angles. 
This factor would be explained in the see-saw mechanism as in the $p=1$ model.

\section{Conclusion} 
\label{sec-concl}

In this work, 
we constructed models to explain the hierarchical structures 
of both quark and lepton sectors under the modular flavor symmetry $\Gamma^\pr_4$, 
which is isomorphic to $S_4^\pr$. 
The hierarchical structure is realized by the FN mechanism 
due to the residual $\mathbbm{Z}_4^T$ symmetry, 
where the small parameter is given by $\eps(\tau)$ having the charge one 
under $\mathbbm{Z}_4^T$. 
In addition, another hierarchical structure 
is induced from the powers of $2\Ita$, 
originated from the canonical normalization depending on the modular weights. 
Unlike the model proposed in our previous paper~\cite{Abe:2023ilq}, 
we consider the case in which the non-hatted and hatted representations 
appear in the same type of fermions, as in $u^c$ of Eq.~\eqref{eq-creps}. 
This generalization allows us to explain the fermion hierarchies 
with the $\order{1}$ coefficients by the $S_4^\pr$ symmetry alone. 
In fact, the ratio of the smallest and largest coefficients are 8.6 (7.6) 
in the first (second) model.  
It is interesting that the small deviations 
from the hierarchical structure by $\eps$ 
in Eqs.~\eqref{eq-texture} and~\eqref{eq-texmix} 
are resolved by the powers of $2\Ita$ 
and the numerical factors in the modular forms.

The lepton sector is studied in addition to the quark sector, 
where we assume that the Majorana neutrino masses are realized 
from the Weinberg operator. 
The weak doublet lepton $L$, including the neutrinos, 
should have a common charge under the $\mathbbm{Z}^T_4$ symmetry, 
so that the lepton mixing angles are $\order{1}$. 
This indicates that the three weak doublet leptons are the same singlet representation. 
The small hierarchy is induced from the powers of $2\Ita$, 
since the weights of $L$ may be different for three massive charged leptons. 
This naturally explains the smallness of the angle $s_{13}^2$, 
while it also predicts small angle $s_{12}^2$, 
see Eqs.~\eqref{eq-estI} and~\eqref{eq-estII}. 
These small differences are simply resolved by the difference 
in the $\order{1}$ coefficients.

\section*{Acknowledgment}
The work of J.K. is supported in part by the Institute for Basic Science (IBS-R018-D1). 
This work is supported in part by he Grant-in-Aid for Scientific Research from the
Ministry of Education, Science, Sports and Culture (MEXT), Japan 
No.\ 22K03601~(T.H.) and 18K13534~(J.K.). 
The work of Y.A. is supported by JSPS Overseas Research Fellowships.

\afterpage{\clearpage} 
\appendix 
\section{Modular forms of $\Gamma_4^\pr$} 
\label{app-S4p}

We collect the modular forms used in the model, 
which can be constructed from products of the modular form at $k=1$ in Eq.~\eqref{eq-Y1}.  
The products of the non-trivial singlets are given by 
\begin{align}
\label{eq-prodsgl}
 1^\pr \otimes 1^\pr = \hsg \otimes \hsg^\pr = 1, 
\quad 
\hsg\otimes\hsg = \hsg^\pr \otimes \hsg^\pr = 1^\pr, 
\quad 
1^\pr \otimes \hsg^\pr = \hsg,     
\quad 
1^\pr \otimes \hsg = \hsg^\pr.  
\end{align}
The products of $2$ and $3$ are given by 
\begin{align}
2(u)\otimes 2(v) =&\  
(u_1v_1+u_2v_2)_1\oplus (u_1v_2-u_2v_1)_{1^\pr} \oplus 
\begin{pmatrix}
 u_2 v_2-u_1v_1 \\ u_1v_2 + u_2v_1 
\end{pmatrix}_2, \\ \notag   
2(u)\otimes 3(\phi) =&\  
\begin{pmatrix}
-2u_1 \phi_1  \\ 
u_1\phi_2 - \sqrt{3} u_2 \phi_3 \\ 
u_1\phi_3 - \sqrt{3} u_2 \phi_2 \\ 
\end{pmatrix}_3
\oplus
\begin{pmatrix}
-2u_2 \phi_1 \\ 
u_2\phi_2 + \sqrt{3} u_1 \phi_3 \\ 
u_2\phi_3 + \sqrt{3} u_1 \phi_2 \\ 
\end{pmatrix}_{3^\pr}, 
\\ \notag 
3(\phi)\otimes 3(\psi) =&\ 
\left(\phi_1\psi_1+\phi_2\psi_3 +\phi_3\psi_2 \right)_1 
\oplus 
\begin{pmatrix}
 2\phi_1\psi_1-\phi_2\psi_3-\phi_3\psi_2 \\ 
 \sqrt{3}\left(\phi_2\psi_2+\phi_3\psi_3\right)
\end{pmatrix}_2 
\\ \notag 
&\quad \oplus
\begin{pmatrix}
\phi_2\psi_2-\phi_3\psi_3 \\ 
-\phi_3\psi_1-\phi_1\psi_3 \\   
 \phi_1\psi_2+\phi_2\psi_1 
\end{pmatrix}_3 
\oplus
\begin{pmatrix}
\phi_2\psi_3-\phi_3\psi_2 \\ 
\phi_1\psi_2-\phi_2\psi_1 \\
\phi_3\psi_1-\phi_1\psi_3
\end{pmatrix}_{3^\pr}.  
\end{align}
The products of the other representations with prime and/or hat 
are calculated in the same manner as the singlets in Eq.~\eqref{eq-prodsgl}, 
with noting that primed doublets, $2^\pr$ and $\hdb^\pr$ denoted 
by $(u_1, u_2)_{\mathbf{2}^\pr}$, 
should be understood as $(u_2, -u_1)_{\mathbf{2}}$, where $\mathbf{2} = 2, \hdb$.

The modular forms used in the models are given by
\begin{align}
 Y^{(3)}_{\htr} =&\   
\begin{pmatrix}
2\sqrt{2} \eps\theta \left(\eps^4 +  \theta^4 \right) \\ 
 -\eps^2(\eps^4 - 5 \theta^4) \\ 
- \theta^2( 5 \eps^4 - \theta^4 ) 
\end{pmatrix},  
& \quad & \\ \notag 
Y^{(4)}_{1} =&\ \eps^8+14\eps^4\theta^4+\theta^8,
& \quad  
 Y^{(4)}_{3} =&\ 
\eps\theta\left(\eps^4-\theta^4\right)  
\begin{pmatrix}
 -\sqrt{2}\eps\theta \\  -\eps^2 \\ \theta^2 
\end{pmatrix},  \\ \notag 
Y^{1(5)}_{\htr} =&\ 
 \eps\theta(\eps^4-\theta^4)
\begin{pmatrix}
\eps^4-\theta^4 \\ 
2\sqrt{2} \eps\theta^{3} \\
2\sqrt{2} \eps^3\theta 
\end{pmatrix}, 
&\quad 
Y^{2(5)}_{\htr} =&\ 
\begin{pmatrix}
16\sqrt{2}\epth{5}{5} \\ 
\eps^2(\eps^{8} + 10 \epth{4}{4}+ 5 \theta^{8}) \\
-\theta^2(5\eps^{8}+ 10\epth{4}{4} +\theta^{8}) 
\end{pmatrix}, \\ \notag   
 Y^{(6)}_1 =&\ 
(\eps^4+\theta^4)(\eps^{8}-34\epth{4}{4}+\theta^8),  
&\quad 
 Y^{(6)}_3 =&\ \eps\theta \left(\eps^4-\theta^4\right)
\begin{pmatrix}
-2\sqrt{2} \eps\theta(\eps^4+\theta^4) \\ 
\eps^2 \left(\eps^4-5\theta^4\right) \\
\theta^2 \left(5\eps^4-\theta^4\right)  
\end{pmatrix}, \\ \notag  
Y^{1(7)}_{\htr} =&\ 
\begin{pmatrix}
 32\sqrt{2} \epth{5}{5}\left(\eps^{4}+\theta^{4}\right) \\ 
 -\eps^2(\eps^{12}-19\epth{8}{4} -45\epth{4}{8} - \theta^{12}) \\ 
 -\theta^2(\eps^{12} + 45\epth{8}{4} +19\epth{4}{8} - \theta^{12}) 
\end{pmatrix}, 
&\quad 
Y^{2(7)}_{\htr} =&\ \eps\theta(\eps^4-\theta^4)
\begin{pmatrix}
\eps^8-\theta^8 \\ 
-\sqrt{2} \eps\theta^3 (7\eps^4+\theta^4) \\ 
-\sqrt{2} \eps^3\theta (\eps^4+7\theta^4) \\ 
\end{pmatrix},  \\ \notag 
 Y^{(8)}_1 =&\  (\eps^8 + 14\epth{4}{4}+\theta^8)^2, 
&\quad & \\ \notag 
 Y^{1(8)}_3 =&\  \eps\theta(\eps^4-\theta^4) 
\begin{pmatrix}
 16\sqrt{2}\epth{5}{5} \\ 
 \eps^2\left(\eps^8+10\epth{4}{4}+5\theta^8 \right)  \\
-\theta^2\left(5\eps^8+10\epth{4}{4}+\theta^8\right) 
\end{pmatrix}, 
\quad&  
 Y^{2(8)}_3 =&\  \epth{2}{2}(\eps^4-\theta^4)^2 
\begin{pmatrix}
\eps^4-\theta^4 \\ 
2\sqrt{2} \eps \theta^3 \\ 
2\sqrt{2} \eps^3 \theta \\  
\end{pmatrix}.  
\end{align}
These are normalized 
such that absolute value of the coefficient of the largest element at $\eps \ll 1$ 
is unity.

{\small
\bibliography{ref_modular} 
\bibliographystyle{JHEP} 
}

\end{document}